\documentclass[%
nofootinbib,
 amsmath,amssymb,
 aps,
 prl,
 twocolumn,
 superscriptaddress
]{revtex4-1}

\usepackage{dcolumn}

\allowdisplaybreaks

\usepackage[dvipdfmx]{graphicx}
\usepackage{latexsym}
\usepackage{amsfonts}
\usepackage{amssymb}
\usepackage{amsmath}
\usepackage{bm}
\usepackage{mathrsfs} 

\newcommand{\bp}{{\bm{p}}}
\newcommand{\bq}{{\bm{q}}}

\newcommand{\Lag}{ {\mathscr{L}} }

\newcommand{\M}{ {\mathcal M} }

\newcommand{\cbar}{{ \bar {\rm c} }}

\newcommand{\ext}{ {\rm ext} }

\newcommand{\LL}{ {\ell \bar \ell} }
\newcommand{\ee}{ {e^+ e^-} }
\newcommand{\mm}{ {\mu^+ \mu^-} }

\def\simge{\mathrel{%
   \rlap{\raise 0.511ex \hbox{$>$}}{\lower 0.511ex \hbox{$\sim$}}}}
\def\simle{\mathrel{
   \rlap{\raise 0.511ex \hbox{$<$}}{\lower 0.511ex \hbox{$\sim$}}}}
\def\bigs{\mathrel{
   \rlap{\raise 0.531ex \hbox{$>$}}{\lower 0.531ex \hbox{$<$}}}}

\usepackage[normalem]{ulem}  
\usepackage[dvips]{color} 
\newcommand{\com}[1]{{\sf\color[rgb]{0,0,1}{#1}}}

\renewcommand\sout{\bgroup \color{red} \ULdepth=-.5ex \ULset}

\usepackage{slashed}

\begin{document} 

\vspace*{-10mm}
\begin{flushright}
KEK-TH-1637
\end{flushright}
\vspace{-5mm}

\vspace*{-10mm}
\begin{flushright}
\end{flushright}
\vspace{-5mm}

\author{Koichi Hattori}\email{\tt khattori@yonsei.ac.kr}
\affiliation{
Institute of Physics and Applied Physics, 
Yonsei University, Seoul 120-749, Korea
}

\author{Kazunori Itakura }\email{\tt kazunori.itakura@kek.jp}
\affiliation{Theory Center, IPNS, 
High energy accelerator research organization (KEK), 
1-1 Oho, Tsukuba, Ibaraki 305-0801, Japan}
\affiliation{Department of Particle and Nuclear Studies, 
Graduate University for Advanced Studies (SOKENDAI), 
1-1 Oho, Tsukuba, Ibaraki 305-0801, Japan}

\author{Sho Ozaki}\email{\tt sho@rcnp.osaka-u.ac.jp}
\affiliation{
Institute of Physics and Applied Physics, 
Yonsei University, Seoul 120-749, Korea
}


\vspace*{5mm}
\title{Neutral--pion reactions induced by chiral anomaly in strong magnetic fields}


\date{\today}

\begin{abstract}
We investigate decay and production of neutral pions in 
strong magnetic fields. 
In the presence of strong magnetic fields, transition between a neutral pion and a virtual photon becomes possible through the triangle diagram relevant for the chiral anomaly. 
We find that the decay mode of a neutral pion into two photons 
cannot persist in the dominant mode in strong magnetic fields, 
and that decay into a dilepton instead 
dominates over the other modes. 
We also investigate the effects of magnetic fields on prompt virtual 
photons created in ultrarelativistic heavy-ion collisions. There is no 
anisotropy in the spectrum at the stage of creation of prompt 
virtual photons, but 
after traversing the strong magnetic field that is induced perpendicularly 
to the reaction plane, virtual photons turn into neutral pions, 
leading to an anisotropic spectrum of dileptons 
as a feasible signature in the measurement. 
\end{abstract}





\maketitle

Much attention has been 
paid 
to strong magnetic fields in nature and laboratories. 
Especially, magnitudes of magnetic fields are thought to reach 
$\vert B \vert \sim 10^{11}$ T in strongly magnetized neutron stars 
as known as {\it magnetars} \cite{TD}, and $\vert B \vert \sim 10^{13}$ T 
at 
the impact of 
ultrarelativistic heavy--ion collisions 
in 
RHIC and LHC \cite{KMW,estimates,Itakura_PIF}. 
A magnitude of the latter field provides a scale as large as pion mass $|eB| \sim m_\pi^2$, 
implying that 
effects of the strong magnetic field on light hadrons could become as important as strong interaction. 
In this Letter, 
we address 
the 
effects of such extremely strong magnetic fields on 
neutral pion reactions through the chiral anomaly \cite{anomaly,anom_rev}.

\begin{figure}[t]
	\begin{center}
		\includegraphics[width=\hsize]{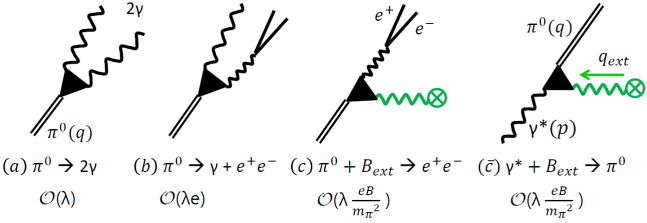}
	\end{center}
\vspace{-0.6cm}
\caption{
Decay and production of a neutral pion through the 
WZW term (filled triangles). 
}
\label{fig:tri}
\vspace{-0.3cm}
\end{figure}

First, we show that the leading decay mode and lifetime of neutral pion change 
as magnitude of an external magnetic field approaches the neutral--pion mass squared, 
$\vert eB \vert \sim m_\pi^2$, 
and/or a propagating pion carries large energy. 
In the strong field limit, a neutral pion dominantly decays into a 
dilepton without being accompanied by any real photon 
in the final state (see Fig.~\ref{fig:tri}(c)).

The inverse process provides a neutral--pion production mechanism (see Fig.~\ref{fig:tri}($\bar {\rm c}$)) 
known as {\it the Primakoff effect}, i.e., a conversion of 
a real photon 
into 
a neutral pion in atomic Coulomb fields \cite{Pri}. 
Remarkably, the Primakoff effect has been 
a standard method in measurement of neutral--pion lifetime \cite{Anom_review}, 
and also has been applied to search for a hypothetical pseudo--scalar particle ``axion'' \cite{axion}. 
We investigate a neutral--pion production mechanism applied to 
the prompt virtual photon from hard parton scatterings 
in ultrarelativistic heavy--ion collisions. 
Since this scattering process takes place in a time scale 
much 
shorter than that of a rapidly decaying strong magnetic 
field \cite{KMW,estimates}, 
prompt photons enjoy much chance to interact with the magnetic fields. 
We show that an oriented production rate with respect to the external 
magnetic field 
gives rise to an anisotropic spectrum of dileptons 
originating from the prompt virtual photons. 
The second Fourier coefficient of the azimuthal 
angle dependence, conventionally called $v_2$, 
will be analytically related to the Fourier component of the spacetime 
profile of the 
magnetic field. 
We will find that energy transfer from a time--dependent magnetic field 
enlarges the final--state phase space, 
leading to a non--vanishing dilepton $v_2$ in a wide kinematical window.  

It has been shown that conservation of the axial vector current $j_5^{\mu} $
is anomalously violated 
\cite{anomaly}: 
\begin{eqnarray}
\partial_\mu j_5^{\mu} =  \frac{ \alpha_{\rm em} }{ 4\pi } F^{\mu\nu} \tilde F_{\mu\nu}
\ \ ,
\label{eq:anom}
\end{eqnarray}
where the fine structure constant is given by $\alpha_{\rm em} = e^2/(4\pi)$ 
with unit electric charge ``$e$". 
This ``anomaly relation" indicates a coupling between neutral--pion field $\pi^0$ 
composing $j_5^{\mu} $ 
and two photon fields, 
and is consistently taken into account as an effective vertex 
called the Wess--Zumino--Witten (WZW) term \cite{WZW}, 
\begin{eqnarray}
\Lag_{\rm WZW} 
&=&
\frac{\lambda}{4} \, \pi^0 \, F^{\mu\nu} \tilde F_{\mu\nu}
\label{eq:WZW}
\ \ .
\end{eqnarray}
A coupling constant $\lambda = ( N_c e^2 ) / ( 12 \pi^2 f_\pi)$ 
is specified by the number of color degrees $N_c = 3$ and pion decay constant $f_\pi = 92.2$ MeV. 
In the presence of an external field, 
we divide the photon field $A^\mu$ into dynamical and external fields as 
$A^\mu = a^\mu + A_{\rm ext}^\mu$, 
and correspondingly the field strength tensor as $F^{\mu\nu} = f^{\mu\nu} + F_{\rm ext}^{\mu\nu}$. 
Substituting these 
into 
the WZW term (\ref{eq:WZW}), 
we obtain not only the conventional vertex with two dynamical photons, 
\begin{eqnarray}
\Lag_{\gamma \gamma} &=& \frac{\lambda}{2} 
\pi^0 \epsilon^{\mu\nu \alpha \beta} ( \partial_\mu a_\nu ) ( \partial_\alpha a_\beta)
\label{eq:Laa}
\ \ ,
\end{eqnarray}
but also a three--point vertex including an external field, 
\begin{eqnarray}
\Lag_{F \gamma} &=& \lambda \pi^0 
( \partial_\mu a_\nu ) \tilde F^{\mu\nu}_{\rm ext}
\label{eq:LaB}
\ \ .
\end{eqnarray}

The Adler--Bardeen theorem tells 
that 
the anomaly relation (\ref{eq:anom}) 
and thus the WZW term (\ref{eq:Laa}) persist in the all--order perturbation theory \cite{AB}, 
which implies that only the triangle diagram without higher--order radiative corrections is relevant. 
In the absence of an external field, 
the WZW term (\ref{eq:Laa}) describes the 
decay of $\pi^0$ 
into two photons 
(Fig.~\ref{fig:tri} (a)), 
providing a decay width to be, $\Gamma_{2\gamma} = \lambda^2 m_\pi^3 / (64 \pi  ) = 7.62 $ eV 
in excellent agreement with measurement: 
$\Gamma_{2\gamma}^{\rm exp} = 7.74 \pm 0.37$ eV. 
The next--to--leading decay mode follows from a QED correction to one of the two photons, 
i.e., the Dalitz decay (Fig.~\ref{fig:tri} (b)), 
without decay into photons 
more than two. 
The Adler--Bardeen theorem 
also confirms that a coupling of $\pi^0$ 
to an external 
field is determined by 
the vertex (\ref{eq:LaB}) 
\cite{NQED}. 

Focusing on the vertex (\ref{eq:LaB}), 
we find a decay mode possible only in the presence of strong external fields. 
We shall consider the case of magnetic fields below. 
As depicted in Fig.~\ref{fig:tri} (c), a neutral pion 
couples to an electromagnetic current 
through the vertex (\ref{eq:LaB}), 
corresponding to 
a dilepton in the final state. 
While the amplitude of this process, $\pi^0(q) + B \rightarrow \gamma^\ast \rightarrow e^+ e^-$, 
is suppressed compared to decay mode (a) 
by an order in QED coupling constant, 
$\mathcal O (\lambda \cdot eB/m_\pi^2)$, 
with an inverse pion mass square from the virtual--photon propagator, 
a large value 
of a strong external magnetic field compensates the suppression 
when the magnitude becomes strong beyond the ``critical field'' defined by $B^c_\pi = m_\pi^2/e$. 
Therefore, {\it dilepton mode (c) overwhelms the Dalitz decay (b) and two--photon mode (a) 
as the magnetic field becomes strong.}

With the vertex (\ref{eq:LaB}), 
a spin--summed decay rate for the decay process (Fig.~\ref{fig:tri}~(c)) 
is expressed as 
\begin{eqnarray}
\hspace{-0.3cm}
\Gamma_{Be^+e^-} &=&
\frac{ \lambda^2}{2\omega_\pi} L^{\mu\nu}(q) 
(q_\alpha i D_{\mu\beta} \tilde F_{\rm ext}^{\alpha\beta} )
(q_\rho i D_{\nu\sigma} \tilde F_{\rm ext}^{\rho\sigma} )^\dagger
\label{eq:Gam0}
\  ,
\end{eqnarray}
where the ``lepton tensor" $L^{\mu\nu}(q)$ provides a 
squared amplitude of the process, $\gamma^\ast \rightarrow e^+e^-$, 
which is related to an imaginary part of the photon vacuum polarization tensor, 
$L^{\mu\nu}(q) = 2 \; {\rm Im} \, \Pi^{\mu\nu} (q) $. 
For the photon propagator $D^{\mu\nu} (q)$ and the imaginary part of the vacuum polarization tensor, 
we incorporate the lowest--order contribution in the order of the QED coupling and the magnetic field: 
namely, the free photon propagator (in arbitrary gauge) 
and one--loop vacuum polarization tensor in the ordinary vacuum. 
Without momentum transfer from a constant magnetic field, 
a pion with $q^2=m_\pi^2$ 
can decay into $e^+e^-$ but not $\mu^+ \mu^-$, 
so that this lepton mass 
corresponds to 
electron mass below.

Inserting the free photon propagator and the one--loop 
vacuum polarization tensor 
into the decay rate (\ref{eq:Gam0}), one finds that 
the gauge--dependent terms in the 
propagators 
drop in contracting with the transverse projection operator in 
$L^{\mu\nu}$, 
and thus the decay rate is obviously gauge invariant. 
With $q^2 = m_\pi^2$, the decay rate is obtained as 
\begin{eqnarray}
\hspace{-0.4cm}
\Gamma_{Be^+e^-} &=&
%
\frac{ q^2  q_\parallel^2 }{ 12 \pi \omega_\pi } \left ( \lambda \frac{eB}{q^2} \right ) ^2
\!\! \left(1+\frac{2m^2}{q^2} \right) \!\! \sqrt{ 1 - \frac{4m^2}{q^2} } 
\label{eq:G_Bee}
\ ,
\end{eqnarray}
where a squared momentum is denoted as 
$q_\parallel^2 = \omega_\pi^2 - q_{\bm b}^2 
= \omega_\pi^2 - \vert \bq \vert^2 \cos ^2 \theta $ 
with $q_{\bm b}$ and $\theta$ being a spatial component of the momentum parallel to the external field 
and an angle measured from the direction of the field, respectively. 
This angle dependence gives rise to an elliptically--anisotropic production rate, 
and the decay rate grows quadratically with respect to $B$ by the 
factor 
$(\lambda\cdot eB/m_\pi^2)^2=(\lambda \cdot B / B_\pi^c)^2$ 
as mentioned above.

\begin{figure}
\vspace{-0.2cm}
  \begin{center}
   \includegraphics[width=\hsize]{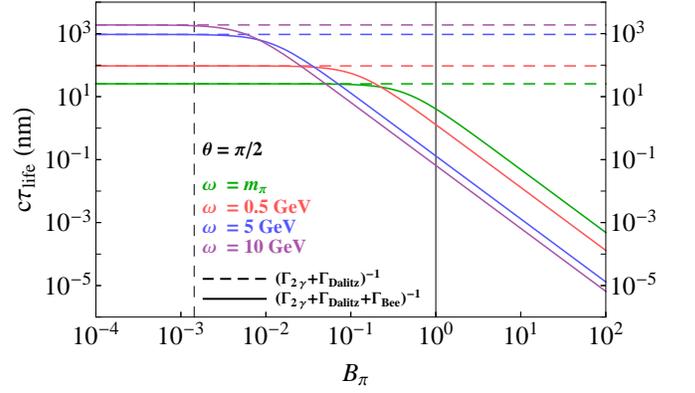}
  \end{center}
\vspace{-0.7cm}
\caption{
Mean lifetime against magnetic--field--strength ($B_\pi = B/B^c_\pi$): 
While dashed lines include contributions from 
decay modes (a) and (b), solid lines include (c) in addition to the other two. 
Pion energy is distinguished by colors. }
\label{fig:life_B}
\vspace{-0.3cm}
\end{figure}

\begin{figure}
  \begin{center}
   \includegraphics[width=\hsize]{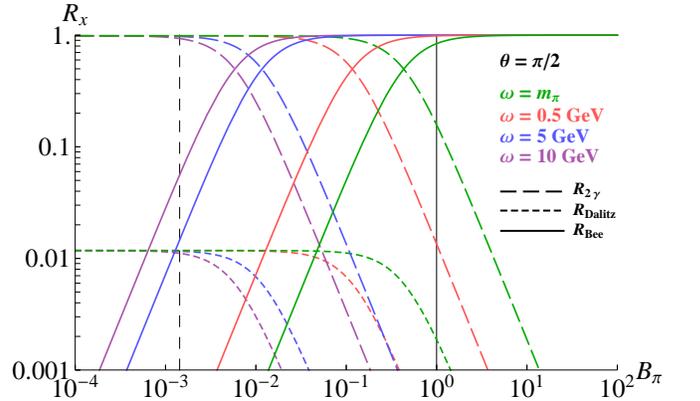}
  \end{center}
\vspace{-0.6cm}
\caption{
Branching ratio against magnetic--field--strength ($B_\pi = B/B^c_\pi$): 
Dashed, dotted and solid lines show branching ratios of modes (a), (b) and (c), respectively. 
}
\label{fig:R_B}
\vspace{-0.35cm}
\end{figure}

Figure~\ref{fig:life_B} shows the lifetime of 
$\pi^0$ in a strong magnetic field, 
obtained as an inverse of the total decay rate, $\tau_{\rm life} = \Gamma_{\rm total}^{-1}$. 
While dashed lines show the lifetime against decay modes (a) and (b) in the ordinary vacuum, 
solid lines indicate lifetime including decay mode (c) as well as the other two, 
which decreases by a few orders 
when the magnetic field becomes strong. 
The decay width in modes (a) and (b) are referred to the measured values 
$\Gamma_{2\gamma}^{\rm rest}  = 7.74  $ eV and $\Gamma_{\rm Dalitz}^{\rm rest}  = 0.0909 $ eV 
with an appropriate kinematical factor providing decay widths in arbitrary Lorentz frame, 
$\Gamma_{2\gamma, \, {\rm Dalitz}} (\omega_\pi) = 
(m_\pi/\omega_\pi) \cdot \Gamma_{2\gamma, \, {\rm Dalitz}} ^{\rm rest}  $. 
As this kinematical factor represents the relativistic time delay, 
a fast moving pion in a weak field has 
larger lifetime than that of a rest pion. 
However, Fig.~\ref{fig:life_B} shows that 
a fast moving pion decays faster 
than a rest pion in strong magnetic fields 
because of an 
enhancement of the decay rate (\ref{eq:G_Bee}) by a factor, $q_\parallel^2$.


To show the dominant decay mode in the presence of strong magnetic fields, 
we define a 
branching ratio by 
$R_x = \Gamma_x / (\Gamma_{2\gamma} + \Gamma_{\rm Dalitz} + \Gamma_{Be^+e^-})$ 
for ``$x$"--mode among (a), (b) and (c). 
Magnetic--field--strength dependence of $R_x$ 
is shown in Fig.~\ref{fig:R_B}. 
Decay mode (c) overwhelms the other two modes 
not only in the strong field limit, 
but also in relatively weak field if pions carry large energy. 
Dashed vertical line shows a 
field strength of the order of 
$B_\pi \sim 100 \cdot (m_e^2/m_\pi^2)$ which are thought to accompany the magnetars.

Very recently, pion spectrum in cosmic rays was constructed from two photons \cite{SNR}, 
of which origin is attributed to energetic collisions of 
highly 
accelerated protons 
in the region of supernova remnants. 
In such radical events in stars, strong magnetic fields extending in macroscopic scales 
may act energetic neutral pions to modify the lifetime and branching ratio, 
and in turn successive processes of energetic nuclear reactions.



One would expect that the strong magnetic fields created in ultrarelativistic heavy--ion collisions 
also provide any signature of the interaction through the anomaly relation. 
Indeed, 
it is proposed that photon production is possible in strong magnetic fields by 
conversions from 
a dilatation current through conformal anomaly \cite{BKS} 
and an axial vector current through chiral anomaly \cite{FM}. 
However, it should be elaborated whether 
light--meson components in those currents 
could interact with the rapidly decaying magnetic field, 
because formation of light mesons would take a longer time than the lifetime of the magnetic field. 
This also holds in the present case. 
Neutral pions are 
expected to be formed 
in a time scale of the order of its mean lifetime 
that is much larger than the lifetime of the magnetic field, 
even if we take into account the short lifetime due to 
the magnetic field shown in Fig.~\ref{fig:life_B}. 
Therefore, we do not expect that significant amount of neutral 
pions decay just after the collisions. 
However, it opens another possibility if we consider the inverse 
process, namely, 
a neutral pion production from a prompt virtual photon. 
In the rest of this Letter, we propose that $v_2$ of the dilepton originating from the prompt virtual photon 
emerges as a reflection of the oriented neutral--pion production 
in the strong magnetic field though the WZW term (\ref{eq:LaB}) (see Fig.~\ref{fig:tri}~($\cbar$)).

First, recall that a cross section of the dilepton production 
from a prompt virtual photon in a nucleon+nucleon collision 
is expressed 
as (see Appendix B in Ref.~\cite{dilepton})
\begin{eqnarray}
\frac{d\sigma^{\LL}_{NN} }{ d^4p } &=& 
\frac{\alpha_{\rm em}}{3\pi} f(m_{\LL}^2)
\frac{d\sigma^{\gamma^\ast}_{NN} }{ d^4p } \, ,
\label{eq:CSee}
\end{eqnarray}
where 
$f(m_{\LL}^2) = \sqrt{ 1- 4m^2/m_{\LL}^2} \, (1+2m^2/m_{\LL}^2) / m_{\LL}^2$ 
and $p^2 = m_{\LL}^2$ is 
an invariant mass of a dilepton. 
The cross section of 
virtual photon production $d\sigma^{\gamma^\ast}_{NN} / d^4p  $ is 
given 
by perturbative QCD calculation \cite{pQCD}. 
Note that the final--state momenta of leptons 
are integrated out in Eq.~(\ref{eq:CSee}), 
and thus this expression provides an integrated cross section 
at a given virtual photon momentum, $p$.
By Glauber--modeling of high--energy collisions, 
total yield in a heavy--ion collision event is obtained by scaling 
the cross section in a proton+proton collision 
with the collision geometry encoded in the overlap function $T_{AB}(b)$ between nuclei A and B 
at impact parameter $b$ \cite{Gla}, resulting in a simple scaling formula, 
$dN^{\LL}_{AB} / d^4q = T_{AB} (b) \cdot d\sigma^{\LL}_{AB} / d^4q $. 

In case of the pion production from a virtual photon in Fig.~\ref{fig:tri}~($\cbar$), 
a squared S--matrix element with the WZW term (\ref{eq:LaB}) is straightforwardly calculated as 
\begin{eqnarray}
\frac{d\sigma^{\pi}_{NN} }{ d^4p } &=& \frac{\lambda^2}{3\pi}
\frac{p_\parallel^2 }{ m_{\LL}^4 } I (p) \frac{d\sigma^{\gamma^\ast}_{NN} }{ d^4p } 
\ \ ,
\label{eq:CSpi}
\\
I(p)  & \equiv& 
\frac{1}{V_4} \int \!\! \frac{d^4 q}{ (2\pi)^4 } 2\pi \delta(q^2 - m_\pi^2) \tilde B^2 (q-p)
\ \ ,
\label{eq:int}
\end{eqnarray}
where the final--state pion phase space is integrated out with respect to $q$, 
so that Eq.~(\ref{eq:CSpi}) provides an integrated cross section of the pion production 
from virtual photons carrying a given momentum $p$. 
In integral (\ref{eq:int}), Fourier component of an external magnetic field is denoted as 
$\tilde B (q_\ext) \!\!=\!\! \int \! B(x) \, e^{i q_\ext x} d^4x$. 
Note that dependences of the cross section on an angle and the magnetic field strength 
follow from a Lorentz contraction, 
$p_\mu \tilde F^{\mu\nu} \tilde F_{\nu \sigma} q^\sigma \!=\! p_\parallel^2  \cdot \tilde B^2 (q_\ext) $, 
and that an angle dependence measured from the reaction plane is thus found to be 
$p_\parallel^2 = (p^0)^2 - p_{\bm b}^2 = a_0 + a_2 \cdot \cos 2\phi$ with 
$ a_0 = \vert \bp \vert^2 /2 + m_{\LL}^2 \ $ and $ a_2 = \vert \bp \vert^2 /2$. 

Some of the virtual photons created at the hard collision are 
converted into neutral pions 
in the presence of strong magnetic field.
Since the conversion rate depends on the angle, it effectively gives rise to 
an anisotropy in the spectrum of dileptons. 
By using the cross sections (\ref{eq:CSee}) and (\ref{eq:CSpi}), 
a reduced amount of dilepton yield in the presence of the magnetic field 
may be given by 
$
d \tilde N ^{\LL}_{AB} / d^4p  \equiv dN^{\LL}_{AB} / d^4p   - dN^{\pi}_{AB} / d^4p 
= w^{\LL} \, ( 1 + 2 v_2^{\LL} \cos 2 \phi )
$
where isotropic and elliptically--anisotropic components are expressed as 
\begin{eqnarray}
\hspace{-0.5cm}
w^{\LL} &=& 
\frac{ T_{AB} }{ 3\pi } \cdot \frac{d\sigma^{\gamma^\ast}_{NN} }{ d^4p } 
\left[ \alpha f(m_{\LL}^2) - a_0 \left( \frac{\lambda}{m_{\LL}^2} \right)^2 \!\!\! I(p) \right]  
,
\label{eq:wee}
\\
\hspace{-0.5cm}
v_2^{\LL} &=&
\frac{ - \frac{ 1 }{ 2 } a_2 \left( \frac{\lambda}{m_{\LL}^2} \right)^2 \!\! I(p) }
{  \alpha f(m_{\LL}^2) - a_0 \left( \frac{\lambda}{m_{\LL}^2} \right)^2 \!\! I(p) }
\label{eq:v2ee}
\ \ .
\end{eqnarray}
In the above, we find 
{\it negative $v_2^\LL$ for dilepton production}, 
because the neutral pion production mechanism 
works more efficiently in perpendicular to the magnetic field as inferred 
from $q_\parallel^2$--dependence in Eq.~(\ref{eq:CSpi}).

We capture the magnetic field created in peripheral heavy--ion collisions 
as a time--dependent, but spatially homogeneous, field 
oriented in perpendicular to the reaction plane ($B_y \not = 0 , \ B_{x,z} = 0$). 
Motivated by the Li\'enard--Wichert potential, 
the non--vanishing component is simply modeled as 
$e B (t) = \kappa \, / ( t^2 \sinh^2 y_b + \zeta^2 )^{3/2} $. 
This mimics profiles obtained in numerical 
simulations \cite{estimates} if we take 
a spatial scale $\zeta$, magnitude of the magnetic field $\kappa$ and 
beam rapidity $y_b$ to be $\zeta = 1$ fm, $\kappa = m_\pi^2 \ (10m_\pi^2)$ 
and $y_b = 5.36 \ (7.98) $ 
for nucleus--nucleus collisions at $\sqrt{s} = 0.2 \ (2.76)$ TeV, respectively. 
It is necessary for investigating impact--parameter dependence 
to take into account a charge distribution in nuclei by more sophisticated analyses, 
and thus we do not go into this point in the present Letter. 
Fourier component of the magnetic field, 
$e \tilde B (\omega_\ext) = (2\pi)^3 \delta^{(3)} (\bq_\ext) \cdot e \mathcal B (\omega_\ext)$, 
is then analytically obtained as 
$
e \mathcal B (\omega_\ext) = 
\sqrt{ 2 / \pi } \cdot
\kappa \, \vert  \omega_\ext \vert / ( \zeta \sinh^2 y_b )
\cdot K_1 \left( \frac{ \zeta \, \vert \omega_\ext\vert  }{ \sinh y_b } \right)
$
with the help of modified Bessel function of the first kind, $K_n(x)$. 
Inserting 
this into Eq.~(\ref{eq:int}), 
the integral is carried out as 
\begin{eqnarray}
I (p) = \frac{1}{T} \cdot \frac{ \mathcal B ^2 ( \varepsilon_\pi - \varepsilon_{\LL} ) }{ \varepsilon_\pi }
\ \ ,
\label{eq:Integ}
\end{eqnarray}
where square of delta functions in spatial components 
are understood as a multiplication of the delta function and three--dimensional volume, 
$[ \, (2\pi) ^3 \delta^{(3)}(\bm 0) \, ]^2 =  V_3 \cdot (2\pi)^3 \delta^{(3)}(\bm q_\ext)$, 
and shorthand notations are introduced as, 
$\varepsilon_\pi = \sqrt{\vert \bp \vert^2 + m_\pi^2} $ 
and $\varepsilon_{\LL} = \sqrt{\vert \bp \vert^2 + m_{\LL}^2} $. 
To obtain the cross section (\ref{eq:CSpi}), 
reaction rate has to be averaged over a `time scale $T$' of the reaction in the magnetic field, 
which is here taken to be of the order of lifetime of the rapidly decaying magnetic field, $T=0.5$ fm/$\it c$.

\begin{figure}
\vspace{-0.5cm}
  \begin{center}
   \includegraphics[width=\hsize]{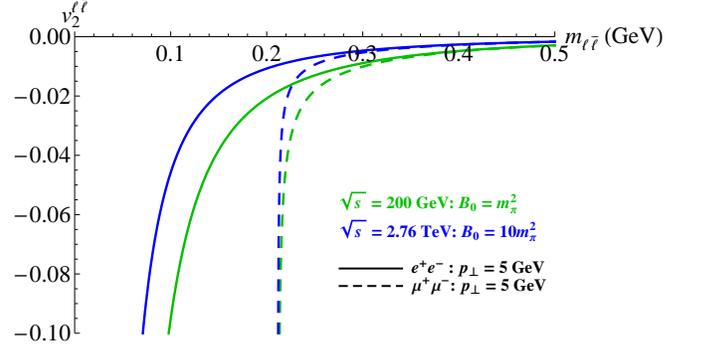}
  \end{center}
\vspace{-0.6cm}
\caption{Dilepton $v_2$ against invariant mass: 
solid and dashed lines show $v_2$ of $e^+e^-$ and $\mu^+\mu^-$ pairs, respectively. 
Profiles of magnetic fields are referred to colors indicated in the legend. 
}
\label{fig:v2_m}
\vspace{-0.4cm}
\end{figure}

\begin{figure}
  \begin{center}
   \includegraphics[width=\hsize]{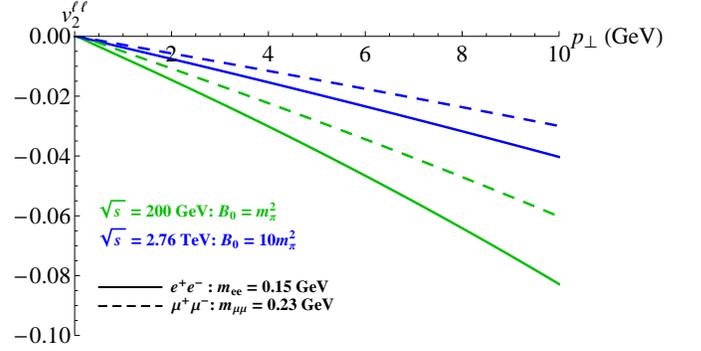}
  \end{center}
\vspace{-0.6cm}
\caption{Dilepton $v_2$ against transverse momentum: 
solid and dashed lines show $v_2$ of $e^+e^-$ and $\mu^+\mu^-$ pairs, respectively. 
}
\label{fig:v2_pt}
\vspace{-0.5cm}
\end{figure}

Figures \ref{fig:v2_m} and \ref{fig:v2_pt} show dilepton $v_2$ at mid--rapidity 
with respect to the invariant mass $m_\LL $ and transverse momentum $p_\perp $, respectively. 
Solid (dashed) line shows a 
result for $\ee$ ($\mm$) pair. 
In Fig.~\ref{fig:v2_m}, the anisotropy becomes large in low invariant mass region, 
since, as seen in Eq.~(\ref{eq:CSpi}), 
pion production is enhanced in this region with less suppression 
by the quartic factor $(m_\LL)^{-4}$ from the virtual photon propagator. 
An energy transfer from the time--dependent magnetic field 
stretches the kinematical window from a single point, $m_\LL = m_\pi$. 
We find in Fig.~\ref{fig:v2_pt} that 
the anisotropy becomes large in high transverse momentum region. 
The reason is twofold. 
First, anisotropy of the longitudinal--momentum square $p_\parallel^2$ in Eq.~(\ref{eq:CSpi}) becomes stronger, 
providing a large $a_2$ in Eq.~(\ref{eq:v2ee}). 
The other depends on profiles of time--dependent magnetic field. 
Noticing that a large transverse momentum provides a small difference 
$\epsilon_\pi - \epsilon_\LL \sim 0 $ in Eq.~(\ref{eq:Integ}), 
high momentum region reflects a value around 
the origin, $\mathcal B ( 0 ) $. 
The Fourier component in the present model takes the largest value at the origin, 
and thus provides a large anisotropy. 
This would hold more generally for a wide variety of the profiles.

In summary, 
we investigated both decay and production of neutral pions in strong magnetic fields. 
These reactions become possible as conversions between neutral pions and virtual photons in external fields. 
The dominant {\it $\pi^0$ decay} into a dilepton provides a large decay width and thus short lifetime of $\pi^0$, 
when the magnetic field strength is large and/or the pion energy is large. 
The decay width is large for pions propagating in perpendicular to the magnetic field, 
since the conversion is the most efficient in this direction. 
This angle dependence of the conversions, 
when applied to {\it $\pi^0$ production} from the prompt virtual photons 
in ultrarelativistic heavy--ion collisions, 
gives rise to an origin of an anisotropic dilepton spectrum totally different from hydrodynamic flow. 
The magnitude of the anisotropy could be as large as, or even larger than, 
that of the thermal dilepton emitted in the quark--gluon plasma phase \cite{dilepton_v2}. 
The negative $v_2$ of dileptons would be more suitable for measurement 
rather than the positive $v_2$ of neutral pions, 
due to a huge background in the pion spectrum and an energy loss in the hot matter. 
We will further pursue a possibility if the effect of the strong magnetic field is 
indeed observed in ultrarelativistic heavy--ion collisions. 
This would require more detailed investigations 
including estimate of a signal significance, i.e., 
the isotropic part (\ref{eq:wee}) combined with a virtual--photon yield from perturbative QCD. 



\section*{Acknowledgements}
This work was supported in part by the Korean Research Foundation under 
Grant Nos. KRF-2011-0020333 and KRF-2011-0030621, 
and also in part by ``The Center for the Promotion of Integrated Sciences (CPIS)" of Sokendai.

\end{document}